\documentclass[journal,12pt,onecolumn,draftclsnofoot,]{IEEEtran}
\IEEEoverridecommandlockouts
\pdfoutput=1
\usepackage{cite}
\usepackage{textcomp}
\usepackage{amsmath,amssymb,amsfonts}
   \usepackage[ruled]{algorithm2e}
   \usepackage{algorithmicx}

\UseRawInputEncoding

\usepackage[american]{babel}%遇单词换行的时候，确保单词的切割是按照音节来而不是随意切割。
\usepackage{microtype}%调整全篇文章（或局部）的字间距，字间距最大调整范围为±1em。可使得某段落不会出现单独一个单词占一行，或文章末尾单独一行文字占一页的不美观情况
\usepackage{graphicx}%插入图片

\usepackage{mathptmx}%把数学字体也改成类似Times的字体。这个字体真的不需要再多说什么了，总之我觉得看久了眼睛会累，但是打印的效果非常稳妥。
\usepackage{amsthm,amsmath,amssymb}

\usepackage{mathrsfs}
\usepackage{mathbbol}%一些特殊的数学公式符号？

\usepackage{caption}
\usepackage{framed}

\usepackage{subfigure}

\usepackage{amsmath}%公式split环境

\usepackage[numbers]{natbib}%使参考文献不带中括号
\setcitestyle{open={},close={}}

\usepackage{stfloats}%双栏中插入单栏的长公式

\usepackage{amsmath}
\usepackage{color}%自定义颜色
\usepackage{mathrsfs}
\usepackage{amsthm}%证明proof环境

\def\BibTeX{{\rm B\kern-.05em{\sc i\kern-.025em b}\kern-.08em
    T\kern-.1667em\lower.7ex\hbox{E}\kern-.125emX}}
\begin{document}

\title{Ergodic Capacity of High Throughput Satellite Systems With Mixed FSO-RF Transmission \\}

\author{Huaicong Kong,
	Min~Lin,
	Zining~Wang,
	Jian Ouyang, 
	and Julian Cheng
\thanks{
This work was supported in part by the Key International Cooperation Research Project under Grant 61720106003, and in part by NUPTSF under Grant NY220111.
 (\emph{Corresponding author: Min Lin.})

H. Kong, M. Lin, Z. Wang and J. Ouyang are with the College of Telecommunications and Information Engineering,
	Nanjing University of Posts and Telecommunications,
	Nanjing, 210003, China	(e-mail: khc\_dream@163.com, linmin@njupt.edu.cn, wzn\_email@163.com, ouyangjian@njupt.edu.cn).

J. Cheng is with the School of Engineering, The University of British
Columbia, Kelowna, BC V1V 1V7, Canada (e-mail: julian.cheng@ubc.ca).

}

}% <-this % stops a space

\maketitle
%\vspace{-20em}
\begin{abstract}
We study a high throughput satellite system, where the feeder link uses free-space optical (FSO) and the user link uses radio frequency (RF) communication.  In particular, we first propose a transmit diversity using Alamouti space time block coding to mitigate the atmospheric turbulence in the feeder link. Then, based on the concept of average virtual signal-to-interference-plus-noise ratio and one-bit feedback, we propose a beamforming algorithm for the user link to maximize the ergodic capacity (EC). Moreover, by assuming that the FSO links follow the  M\'{a}laga  distribution whereas RF links undergo the shadowed-Rician fading, we derive a closed-form EC expression of the considered system. Finally, numerical simulations validate the accuracy of our theoretical analysis, and show that the proposed schemes can achieve higher capacity compared with the reference schemes.
\end{abstract}


\begin{IEEEkeywords}
	High throughput satellite, Beamforming, Ergodic capacity, Mixed FSO-RF transmission.
\end{IEEEkeywords}

% \fi
\IEEEpeerreviewmaketitle

\vspace{-0.2cm}
\section{Introduction}
% no 
\IEEEPARstart
{C}{ompared} with monobeam satellites, high throughput satellites (HTSs), which exploit large amounts of spot beams to provide a larger aggregate capacity and more service flexibility for the users on the earth, have gained attention from both academia and industry [\cite{8746876-1}]-[\cite{9205852-3}].
In such an HTS system, each gateway serves many user beams, whereas each beam is controlled by a single gateway. Consequently, the feeder links require much more capacity than the user links. 
Moreover, it is challenging for the feeder links, currently operating at Ka band, to meet the required bandwidth that increases proportionally with the number of user beams in HTS systems. To address this challenge, the feeder link can adopt free-space optical (FSO) technology owing to its unique advantages, such as unlicensed spectrum, free from interference and high security, over radio frequency (RF) counterpart  [\cite{7553489-4}]-[\cite{8739771-6}]. Many works has focused on the FSO feeder links of HTS systems in the literature.

On the other hand, frequency reuse has been widely used in the user links of HTS systems, where all beams share the same frequency to enhance spectral efficiency of the system. However, this technique introduces inter-beam interference (IBI) due to overlapping side lobes of adjacent beams. The IBI can be mitigated by using precoding or beamforming (BF) techniques in the HTS systems. Traditional methods for mitigating IBI in the user link require exact channel state information (CSI), which is impractical because of the latency and amount of feedback. Recently, a zero-forcing (ZF) BF method has been proposed [\cite{9152013-7}], [\cite{9108615-8}] to maximize the ergodic capacity (EC) of HTS systems, where the locations of each user are exploited when designing BF. 
Although ZF beamforming can mitigate IBI, its performance cannot always be guaranteed due to the limitation of array freedom. This motivates our work presented in this paper.

We here study the performance of the forward link in an HTS system, where the feeder link adopts FSO technology, and the user link, operating at Ka band, serves multiple users simultaneously. Specifically, to mitigate the atmospheric turbulence in the feeder link, we propose an Alamouti space time block coding (STBC) scheme with low complexity. As for the user link, based on the average virtual signal-to-interference-plus-noise ratio (SINR) and one-bit feedback, we propose a BF algorithm to maximize the system EC. Assuming a buffer-aided decode-and-forward (DF) protocol is adopted at the HTS, we derive a closed-form EC expression under the condition that the FSO links experience the M\'{a}laga  fading and RF links follow the shadowed-Rician distribution. Unlike the existing works, which are unsuitable for all regimes of turbulence, we consider a more general model to describe the atmospheric turbulence conditions. In addition, we also consider that the locations of users are available at the HTS, as in the ZF method [\cite{9152013-7}], [\cite{9108615-8}], but our proposed BF algorithm for the user link can achieve higher system capacity, especially at low SNR.

\emph{Notation:} We use $E\left[  \cdot  \right]$, ${( \cdot )^H}$, $\left|  \cdot  \right|$, $\left\|  \cdot  \right\|$, ${\left(  \cdot  \right)_n}$ and ${\mathbb{C}^{M \times N}}$   for the expectation, the Hermitian transpose, the absolute value, the Euclidean norm, the Pochhammer symbol and the complex space, respectively. ${{\mathbf{I}}_N}$ and ${{\mathbf{1}}_K}$  are the identity matrix and all-one vector, respectively. ${J_n}\left( \cdot \right)$ is the ${n}$-th order of first-kind Bessel functions. ${\mathbf{a}} \odot {\mathbf{b}}$  is the Hadamard product. Besides, $G_{p,q}^{m,n}\left[ { \cdot \left|  \cdot  \right.} \right]$, $\Gamma \left(  \cdot  \right)$, ${}_1{F_1}\left( { \cdot ; \cdot ; \cdot } \right)$ and ${\text{Ei}}\left( x \right)$ are the Meijer's G-function, the Gamma function, the confluent hypergeometric function and the exponential integral function, respectively.
\vspace{-0.25cm}
\section{System Model and Problem Formulation}
We consider the forward link of an HTS system, where two gateways communicate with the HTS through FSO feeder link to provide higher capacity, while HTS simultaneously serves $K$ uniformly distributed users via $N$ beams in the user links. Different from the existing works using a single gateway in the feeder link [\cite{9152013-7}], [\cite{9108615-8}], we propose a more general and realistic framework, where two gateways employ the Alamouti STBC scheme to obtain diversity gain without CSI for the feeder link.
In the following, we will describe the channel models of feeder and user links.
\vspace{-0.25cm}
\subsection{Channel Models}
By taking various practical aspects of FSO channels into account, we model the feeder link of the HTS as\footnote[1]{The acquisition, tracking, and pointing system can be employed to address the pointing error problem [4].}[\cite{kaushal2017free-9}]
\vspace{-0.2cm}
\begin{equation}\label{1}
{I_i} = I_i^lI_i^a,i = 1,2
\end{equation}
where $I_i^l = {G_t}{G_r}{\eta _p}{\ell _s}$ and where ${G_t}$ and ${G_r}$  are, respectively, the transmitter and receiver gain; ${\eta _p}$ and ${\ell _s}$  are, respectively, the pointing loss and free-space loss. In addition, the irradiance $I_i^a$  in (1) is assumed to follow the M\'{a}laga  fading, whose probability density function (PDF) is given by [\cite{7859265-10}]
\vspace{-0.2cm}
\begin{equation}\label{2}
{f_{I_i^a}}\left( x \right) = \frac{A}{2}\sum\nolimits_{j = 1}^\beta  {{c_j}{x^{ - 1}}G_{0,2}^{2,0}\left[ {\frac{{\alpha \beta x}}{{{g_0}\beta  + {\Omega ^{'}}}}\left| {\begin{array}{*{20}{c}}
			-  \\
			{\alpha ,j}
			\end{array}} \right.} \right]}
\end{equation}
where
\vspace{-0.5cm}
	\begin{equation}\label{3}
 {\begin{array}{*{20}{c}}
		{A \triangleq \frac{{2{\alpha ^{\frac{\alpha }{2}}}}}{{g_0^{1 + \frac{\alpha }{2}}\Gamma \left( \alpha  \right)}}{{\left( {\frac{{{g_0}\beta }}{{{g_0}\beta  + {\Omega ^{'}}}}} \right)}^{\frac{\alpha }{2} + \beta }},} \\
		{{c_j} \triangleq \left(\!\!\!\! {\begin{array}{*{20}{c}}
				{\beta  \!\!-\! \!1} \\
				{j \!\!-\! \! 1}
				\end{array}} \!\!\!\!\right)\!\!\frac{{{{\left( {{g_0}\beta  + {\Omega ^{'}}} \right)}^{1\! -\! \frac{j}{2}}}}}{{\left( {j - 1} \right)!}}\!\!{{\left(\! {\frac{{{\Omega ^{'}}}}{{{g_0}}}} \!\right)}^{j \!-\! 1}}\!\!{{\left(\! {\frac{\alpha }{\beta }} \!\right)}^{\frac{j}{2}}}\!\!{{\left(\!\! {\frac{{\alpha \beta }}{{{g_0}\beta  + {\Omega ^{'}}}}} \!\!\right)}^{ \!-\! \frac{{\alpha  + j}}{2}}}},
		\end{array}}
	\end{equation}
where  $\alpha $ represents a positive parameter related to the effective number of large-scale cells of the scattering process, $\beta  \in \mathbb{N}$  denotes the amount of fading parameter and  ${g_0} = 2{b_0}\left( {1 - {\rho _0}} \right)$ with $2{b_0}$ being the average power of the total scatter components, and $0 \leqslant {\rho _0} \leqslant 1$  being the amount of the scattering power coupled to the line-of-sight (LoS) component,  ${\Omega ^{'}} = {\Omega _0} + 2{b_0}{\rho _0} + 2\sqrt {2{b_0}\Omega _0 {\rho _0}} \cos \left( {{\phi _A} - {\phi _B}} \right)$ with ${\Omega _0}$  being the average power of the LoS component. Besides, ${\phi _A}$ and ${\phi _B}$ are, respectively, the deterministic angles for the LoS component and the coupled-to-LoS scatter terms.
%In HTS systems, the influence of antenna gain and path loss should be considered to model the RF channel more accurately. 
In HTS systems, the RF channel vector of user link can be written as [\cite{8894851-2}], [\cite{9108615-8}]
\vspace{-0.15cm}
\begin{equation}\label{4}
{{\mathbf{h}}_k} = \sqrt {G_R^{}} {\rho _k} \odot {\mathbf{g}}_k^{1/2} \odot {{\mathbf{\tilde h}}_k},k = 1,2, \cdots ,K
\end{equation}
where $G_R^{}$  denotes the receiver antenna gain at the user, and ${{\mathbf{g}}_k} = {\left[ {{g_{k1}},{g_{k2}}, \ldots ,{g_{kN}}} \right]^T}$ denotes the $N \times 1$  beam gain vector, whose component ${g_{kn}}$  can be expressed as
${g_{kn}} = {g_{\max }}{\left( {\frac{{{J_1}\left( {{u_{kn}}} \right)}}{{2{u_{kn}}}} + 36\frac{{{J_3}\left( {{u_{kn}}} \right)}}{{u_{kn}^3}}} \right)^2}$
where ${g_{\max }}$  is the maximum beam gain, and  ${u_{kn}} = 2.07123\sin {\phi _{kn}}/\sin {\phi _{3{\text{dB}}}}$ where  ${\phi _{kn}}$ is the angle between the $\emph{k}$-th user position and the $\emph{n}$-th beam center with respect to satellite, and  ${\phi _{3{\text{dB}}}}$ is its one-sided half power beam width. 
Besides, we express ${{\mathbf{\tilde h}}_k} = {\left[ {{{\tilde h}_{k1}},{{\tilde h}_{k2}},...,{{\tilde h}_{kN}}} \right]^T}$  with its elements as $
{\tilde h_{kn}} = {\tilde h_\ell }{e^{ - j\frac{{2\pi {f_c}}}{c}{d_{kn}}}}$
where  ${\tilde h_\ell } = \frac{c}{{4\pi {f_c}{d_{kn}}}}$ with  ${f_c}$ denoting the carrier frequency, $c$  being the speed of light, and ${d_{kn}}$  being the distance between the HTS and the $\emph{k}$-th user. In (4),  ${\rho _k}$ refers to the channel fading for the $\emph{k}$-th user, which follows the shadowed-Rician distribution, commonly used for satellite channels, having the PDF [\cite{9108615-8}]
\vspace{-0.15cm}
\begin{equation}\label{5}
\begin{split}
{f_{\left| {{\rho _k}} \right|}}\left( x \right)& = {\left( {\frac{{2{b_k}{m_k}}}{{2{b_k}{m_k} + {\Omega _k}}}} \right)^{{m_k}}}\frac{x}{{{b_k}}}\exp \left( { - \frac{{{x^2}}}{{2{b_k}}}} \right)\\
&\times {}_1{F_1}\left( {{m_k};1;\frac{{{\Omega _k}{x^2}}}{{2{b_k}\left( {2{b_k}{m_k} + {\Omega _k}} \right)}}} \right)
\end{split}
\end{equation}
where ${m_k}$  is the fading severity parameter; $2{b_k}$  and ${\Omega _k}$  are the average power of the multipath component and LoS component, respectively.
\vspace{-0.25cm}
\subsection{Problem Formulation}
The overall communication consists of two phases. In the first phase, the Alamouti STBC scheme is employed, and it can achieve diversity gain without CSI at the transmitter. 
For the FSO feeder link, we consider a dense wavelength division multiplexing system with subcarrier intensity modulation [\cite{9152013-7}]. Thus, the received electrical signals ${{\mathbf{y}}_i} \in {\mathbb{C}^{k \times 1}}$  at the HTS can be expressed as
\vspace{-0.15cm}
\begin{equation}\label{6}
{{\mathbf{y}}_i} = \eta \left( {I_1^2 + I_2^2} \right){{\mathbf{s}}_i} + {{\mathbf{n}}_i},i = 1,2
\end{equation}
where ${{\mathbf{s}}_i}$  denotes symbol vector with $E\left[ {{\mathbf{s}}_i^H{{\mathbf{s}}_i}} \right] = {P_1}$; $\eta $  is the optical-to-electrical conversion coefficient and ${{\mathbf{n}}_i}$  is the additive white Gaussian noise (AWGN) with mean zero and covariance matrix $E\left[ {{{\bf{n}}_i}{\bf{n}}_i^H} \right] = \left( {I_1^2 + I_2^2} \right){N_0}{{\bf{I}}_k}$. From (6), the output instantaneous SNR at the HTS can be expressed as
\vspace{-0.15cm}
\begin{equation}\label{7}
{\gamma _1} = {P_1}{\eta ^2}\left( {I_1^2 + I_2^2} \right)/{N_0}.
\end{equation}

In the second phase, the HTS first decodes its received electrical signal. Then, the recoded signal  ${x_k},k = 1,2, \cdots ,K$ with  $E\left[ {{{\left| {{x_k}} \right|}^2}} \right] = 1$ is sent to the $\emph{k}$-th user with transmit power ${P_{2,k}}$. Moreover, the HTS employs transmitting BF with normalized weight vector ${{\mathbf{w}}_k} \in {\mathbb{C}^{N \times 1}}$, and the received signal at the $\emph{k}$-th user given by
\vspace{-0.15cm}
\begin{equation}\label{8}
{y_{2,k}} = \sqrt {{P_{2,k}}} {\bf{h}}_k^H{{\bf{w}}_k}{x_k} + \sum\nolimits_{j = 1,j \ne k}^K {\sqrt {{P_{2,j}}} {\bf{h}}_k^H{{\bf{w}}_j}{x_j}}  + {n_{2,k}}
\end{equation}
where ${n_{2,k}}$ is the AWGN with mean zero and variance $\sigma _{}^2$. Correspondingly, the instantaneous SINR at the $\emph{k}$-th user is
\vspace{-0.15cm}
\begin{equation}\label{9}
{\gamma _{2,k}} = \frac{{{P_{2,k}}{{\left| {{\mathbf{h}}_k^H{\mathbf{w}}_k^{}} \right|}^2}}}{{\sum\nolimits_{j = 1,j \ne k}^K {{P_{2,j}}{{\left| {{\mathbf{h}}_k^H{\mathbf{w}}_j^{}} \right|}^2}}  + \sigma _{}^2}}.
\end{equation}
According to buffer-aided DF principle, the EC of the forward link is given by [\cite{8970423-11}]
\vspace{-0.15cm}
\begin{equation}\label{10}
C = \min \left( {{C_1},{C_2}} \right)
\end{equation}
where ${C_1} \!\!=\!\! E\left[ {{{\log }_2}\left( {1 \!\!+\!\! {\gamma _1}} \right)} \right]$ and ${C_2}\!\! =\!\! \sum\nolimits_{k = 1}^K{C_{2,k}}$ and where ${C_{2,k}} =  E\left[ {{{\log }_2}\left( {1 + {\gamma _{2,k}}} \right)} \right]$  are, respectively, the EC of the  feeder link and the user link.
To maximize system EC, we find the optimal BF vectors ${\bf{w}}_k^{}$ that maximize EC ${C_2}$, i.e.,
\vspace{-0.15cm}
	\begin{equation}\label{11}
\begin{split}
&\mathop {\max }\limits_{{{\bf{w}}_k}} \sum\limits_{k = 1}^K\!\! {{C_{2,k}} \!\!=\!\! } \sum\limits_{k = 1}^K \!\!{E\!\!\left[\!\! {{{\log }_2}\!\!\left(\!\! {1 \!\!+\!\! \frac{{{P_{2,k}}{{\left| {{\bf{h}}_k^H{\bf{w}}_k^{}} \right|}^2}}}{{\sum\nolimits_{j = 1,j \ne k}^K {{P_{2,j}}{{\left| {{\bf{h}}_k^H{\bf{w}}_j^{}} \right|}^2}} \!\! +\!\! \sigma _{}^2}}} \!\!\right)} \!\!\right]} \\
&{\rm{s}}{\rm{.t}}{\rm{.   }}\quad{\bf{w}}_k^H{\bf{w}}_k^{} = 1.
\end{split}
	\end{equation}
%The analytical expression of ${C_{2,k}}$ is complex making (11) be mathematically intractable.
Using the Jensen's inequality, we can obtain an upper bound of ${C_{2,k}}$ as ${\bar C_{2,k}} = {\log _2}\left( {1 + E\left[ {{\gamma _{2,k}}} \right]} \right)$.
Since directly solving the original problem (11) is intractable, instead we consider the average virtual SINR maximization problem after establishing the connection between
the EC and the average virtual SINR.
The virtual SINR is defined as the ratio between the signal power and a weighted sum of the interference powers plus the noise power [\cite{6130547-12}]
\vspace{-0.15cm}
\begin{equation}\label{12}
{\hat \gamma _{2,k}} \buildrel \Delta \over = \frac{{{P_{2,k}}{{\left| {{\bf{h}}_k^H{\bf{w}}_k^{}} \right|}^2}}}{{\sum\nolimits_{j = 1,j \ne k}^K {{P_{2,k}}{\mu _{k,j}}{{\left| {{\bf{h}}_j^H{\bf{w}}_k^{}} \right|}^2} + \sigma _{}^2} }},
\end{equation}
where ${\mu _{k,j}}$  denotes the weight coefficient to be optimized.
Then, according to the Mullen's inequality [\cite{Mullen-13}], we can obtain an approximate expression for the average virtual SINR  ${\hat \gamma _{2,k}}$ as
\vspace{-0.15cm}
\begin{equation}\label{13}
E\left[ {{{\hat \gamma }_{2,k}}} \right] \approx \frac{{{P_{2,k}}E\left[ {{{\left| {{\bf{h}}_k^H{\bf{w}}_k^{}} \right|}^2}} \right]}}{{\sum\nolimits_{j = 1,j \ne k}^K {{P_{2,k}}{\mu _{k,j}}E\left[ {{{\left| {{\bf{h}}_j^H{\bf{w}}_k^{}} \right|}^2}} \right]}  + \sigma _{}^2}}.
\end{equation}
From (13), we find that BF vectors are independent. Thus, the optimization problem equivalent to (11) can be expressed as
\vspace{-0.2cm}
\begin{equation}\label{14}
\begin{split}
&\mathop {\max }\limits_{{{\bf{w}}_k}, {\mu _{k,j}}} \frac{{{P_{2,k}}E\left[ {{{\left| {{\bf{h}}_k^H{\bf{w}}_k^{}} \right|}^2}} \right]}}{{\sum\nolimits_{j = 1,j \ne k}^K {{P_{2,k}}{\mu _{k,j}}E\left[ {{{\left| {{\bf{h}}_j^H{\bf{w}}_k^{}} \right|}^2}} \right]}  + \sigma _{}^2}}\\
&{\rm{s}}{\rm{.t}}{\rm{.   }}\quad{\bf{w}}_k^H{\bf{w}}_k^{} = 1.
\end{split}
\end{equation}
In the following section, we will find the  BF vectors ${\bf{w}}_k^{}$ and weight coefficients ${\mu _{k,j}}$ for problem (14).
 \vspace{-0.1cm}
\section{Proposed BF Algorithm for the User Link}
Similar to the ZF method [\cite{9152013-7}], [\cite{9108615-8}] under the assumption that LoS is dominant, we also consider that user locations are exploited to design BF vectors without requiring any CSI at the HTS. First, when the weight coefficients ${\mu _{k,j}}$ are fixed, using the Rayleigh quotient and a matrix inverse identity, we obtain the BF vectors ${\bf{w}}_k^{}$ by
\vspace{-0.15cm}
	\begin{equation}\label{15}
	\begin{split}
	{\mathbf{w}}_k^{} &= \arg \mathop {\max }\limits_{{\mathbf{w}}_k^H{\mathbf{w}}_k^{} = 1} \frac{{{P_{2,k}}{{\left| {{{\mathbf{a}}^H}({\phi _k}){\mathbf{w}}_k^{}} \right|}^2}}}{{\sum\nolimits_{j = 1,j \ne k}^K {{P_{2,k}}{\mu _{k,j}}{{\left| {{{\mathbf{a}}^H}({\phi _j}){\mathbf{w}}_k^{}} \right|}^2}}  + {\sigma ^2}}} \\
	&= \frac{{\left(\!\! {{\sigma ^2}{\mathbf{I}} \!\!+\!\! \sum\nolimits_{j = 1,j \ne k}^K {{P_{2,k}}{\mu _{k,j}}{{\mathbf{a}}}({\phi _j}){{\mathbf{a}}^H}({\phi _j})} } \right)\!\!^{-1}{{\mathbf{a}}}({\phi _k})}}{{\left\| {\left(\!\! {{\sigma ^2}{\mathbf{I}}\!\! +\!\! \sum\nolimits_{j = 1,j \ne k}^K {{P_{2,k}}{\mu _{k,j}}{{\mathbf{a}}}({\phi _j}){{\mathbf{a}}^H}({\phi _j})} } \right)\!\!^{-1}{{\mathbf{a}}}({\phi _k})} \right\|}} \\
	\end{split}
	\end{equation}
where  ${\mathbf{a}}({\phi _k}) = \frac{{\sqrt {G_R^{}} c}}{{4\pi {f_c}{d_{kn}}}}{\mathbf{g}}_k^{1/2}$ includes the location and antenna beam pattern of the $\emph{k}$-th user.
 Next, to update the weight coefficient ${\mu _{k,j}}$ in the case of fixed BF vectors, we present the gradient expressions for EC and average virtual SINR maximization problems to identify  that these two gradients have the same zero point.
After some algebraic manipulation, two gradient expressions become

 \vspace{-0.2cm}
	\begin{equation}\label{16}
	\begin{split}
	{\nabla _{{\mathbf{w}}_k^{}}}\left( {{\bar C_2}} \right) &= \frac{2}{{{I_k} + {D_k}}}{\mathbf{a}}({\phi _k}){{\mathbf{a}}^H}({\phi _k}){\mathbf{w}}_k^{} \\
	& - 2\sum\nolimits_{j \ne k}^K {\frac{{{D_j}}}{{{I_j}\left( {{D_j} + {I_j}} \right)}}} {\mathbf{a}}({\phi _j}){{\mathbf{a}}^H}({\phi _j}){\mathbf{w}}_k^{} ,\\
	\end{split}
	\end{equation}
	 \vspace{-0.2cm}
	\begin{equation}\label{17}
	\begin{split}
	&{\nabla _{{\mathbf{w}}_k^{}}}\left( {  {E\left[ {{{\hat \gamma }_{2,k}}} \right]} } \right) = \frac{2}{{{D_k}}}{\mathbf{a}}({\phi _k}){{\mathbf{a}}^H}({\phi _k}){\mathbf{w}}_k^{} \\
	& - 2\sum\nolimits_{j \ne k}^K {\frac{{{\mu _{k,j}}}}{{{\sigma ^2}{\text{ + }}\sum\nolimits_{l \ne k}^K {{\mu _{k,l}}{{\left| {{{\mathbf{a}}^H}({\phi _l}){\mathbf{w}}_k^{}} \right|}^2}} }}} {\mathbf{a}}({\phi _j}){{\mathbf{a}}^H}({\phi _j}){\mathbf{w}}_k^{} ,\\
	\end{split}
	\end{equation}
where ${I_j} = {\sigma ^2}{\text{ + }}\sum\nolimits_{l \ne j}^K {{P_l}{{\left| {{{\mathbf{a}}^H}({\phi _j}){\mathbf{w}}_l^{}} \right|}^2}} $ and ${D_j} = {P_j}{\left| {{{\mathbf{a}}^H}({\phi _j}){\mathbf{w}}_j^{}} \right|^2}$. From (16) and (17), we can observe that two expressions become zero at the same point when
\vspace{-0.2cm}
	\begin{equation}\label{18}
	\frac{{{D_j}\left( {{I_k} + {D_k}} \right)}}{{{D_j}\left( {{I_j} + {D_j}} \right)}} = \frac{{{\mu _{k,j}}{D_k}}}{{{\sigma ^2}{\text{ + }}\sum\nolimits_{l \ne k}^K {{\mu _{k,l}}{{\left| {{{\mathbf{a}}^H}({\phi _l}){\mathbf{w}}_k^{}} \right|}^2}} }},j \ne k.
	\end{equation}
After some manipulation, we can obtain a matrix equation as ${\sigma ^2}{{\mathbf{1}}_{K - 1}} = {{\mathbf{Q}}_k}{{\mathbf{\tilde \mu }}_k}$ where  ${{\mathbf{\tilde \mu }}_k} = \left( {{\mu _{k,1}}, \cdots ,{\mu _{k,k - 1}},{\mu _{k,k + 1}}, \cdots ,{\mu _{k,K}}} \right)$ and the matrix ${{\bf{Q}}_k}$ can be expressed as
\begin{figure*}[ht]%b!
	\begin{equation*}
	\begin{aligned}
	{{\bf{Q}}_k} &= \frac{{{D_k}}}{{{I_k} + {D_k}}}diag\left\{ {\frac{{{D_j}\left( {{I_j} + {D_j}} \right)}}{{{D_j}}},j \ne k} \right\}- \\
&	{{\bf{1}}_{K - 1}}\left( {{{\left| {{{\bf{a}}^H}({\phi _1}){\bf{w}}_k^{}} \right|}^2}, \cdots ,{{\left| {{{\bf{a}}^H}({\phi _{k - 1}}){\bf{w}}_k^{}} \right|}^2},{{\left| {{{\bf{a}}^H}({\phi _{k + 1}}){\bf{w}}_k^{}} \right|}^2}, \cdots ,{{\left| {{{\bf{a}}^H}({\phi _K}){\bf{w}}_k^{}} \right|}^2}} \right).
	\end{aligned}
	\end{equation*}
%	\noindent \rule{\textwidth}{.5pt}%\vskip3pt
\end{figure*}

Therefore, the weight coefficient  ${\mu _{k,j}},j \ne k$ can be updated according to the equation ${{\mathbf{\tilde \mu }}_k} = {\sigma ^2}{\mathbf{Q}}_k^{ - 1}{{\mathbf{1}}_{K - 1}}$, which is termed as step 1. Afterwards, we can compute each BF vector ${\mathbf{w}}_k^{}$ using (15) for the given weight coefficients ${\mu _{k,j}}$, which is termed as step 2. Moreover, since the problem for optimal BF vector is bounded, and has a non-decreasing value with respect to iteration number, both step 1 and step 2 should be operated iteratively until convergence\footnote[2]{Since the derived closed-form EC expression takes a complicated form, it is challenging to show that the exact EC is non-decreasing analytically. Thus, we defer a rigorous convergence analysis to a future work.}, and finally the BF vectors can be obtained. 
	However, the QoS for each user is not guaranteed. To solve this issue, we present a user selection scheme having low complexity. Specifically, instead of providing full CSI, each user sends only one-bit feedback to indicate if its SINR is above or below a predetermined threshold ${\Lambda _{th}}$. Note that only the users who meet the requirement can realize transmission in the user link.
Algorithm 1 summarizes the the proposed BF algorithm based on the average virtual SINR and one-bit feedback.
\vspace{-0.4cm}

\begin{LinesNumbered}
\begin{algorithm}
		\caption{Proposed BF algorithm based on the average virtual SINR and one-bit feedback.}%Ëã·¨Ãû×Ö
		\LinesNumbered %ÒªÇóÏÔÊ¾ÐÐºÅ
		
	Initialize ${\mathbf{U}} = \left\{ {{\text{U}}{{\text{ser}}_1},{\text{U}}{{\text{ser}}_2}, \cdots ,{\text{U}}{{\text{ser}}_K}} \right\}$, where  ${\mathbf{U}}$ refers to the group selected users\;
		Set the stopping criterion $\varepsilon $ and threshold ${\Lambda _{th}}$\;
		
		\While{ ${\mathbf{U}} \ne \emptyset $}{
			Initialize  ${\mathbf{w}}_k^0$,  $k = 1,2, \cdots ,K$ and $t = 0$\;
			\For{$\left\| {{\mathbf{w}}_k^{t + 1} - {\mathbf{w}}_k^t} \right\| > \varepsilon $}{
				Compute $D_k^{t + 1}, I_k^{t + 1}, {\mathbf{Q}}_k^{t + 1}$, $k = 1,2, \cdots ,K$\;
				Update the weight coefficients according to  ${\mathbf{\tilde \mu }}_k^{t + 1} = {\sigma ^2}{\left( {{\mathbf{Q}}_k^{t + 1}} \right)^{ - 1}}{{\mathbf{1}}_{K - 1}}$, $k = 1,2, \cdots ,K$\;
				Compute the BF vectors ${\mathbf{w}}_k^{t + 1}$  according to (15)\;
			}
			Calculate ${\gamma _{2,k}}$ according to (9)\;
			\eIf{${\gamma _{2,k}} < {\Lambda _{th}}$}{
				${\mathbf{U}} = {\mathbf{U}}\backslash \left\{ {{\text{U}}{{\text{ser}}_k}} \right\}$, $k = 1,2, \cdots ,K$\;}{
				${{\mathbf{w}}_k^*} \leftarrow {{\mathbf{w}}_k^{t + 1}}$\;
			}
			
		}
	\end{algorithm}
\end{LinesNumbered}

\vspace{-0.6cm}
\section{ Ergodic Capacity}
In this section, we first provide the statistical characterizations of the FSO and RF channels, and then derive the closed-form expression of the ergodic capacity for the proposed schemes.
\vspace{-0.2cm}
\subsection{Statistical Characterization of FSO and RF Channels}
Denoting the average SNR for the feeder link by  ${\bar \gamma _{1,i}} = {P_1}{\left( {\eta I_i^\ell } \right)^2}/{N_0}$ and according to (2), the moment generating function (MGF) of  ${\gamma _{1,i}} = {\bar \gamma _{1,i}}{\left( {I_i^a} \right)^2}$ is given by
\vspace{-0.15cm}
	\begin{equation}\label{19}
	\setlength{\abovedisplayskip}{3pt}
	\setlength{\belowdisplayskip}{3pt}
	\begin{gathered}
	{M_{{\gamma _{1,i}}}}\left( s \right) = \frac{A}{4}\sum\nolimits_{j = 1}^\beta  {{c_j}}    \hfill \\
\times	\underbrace {\int_0^\infty  {{x^{ - 1}}\exp \left( { - sx} \right)} G_{0,2}^{2,0}\left[ {\frac{{\alpha \beta }}{{{g_0}\beta  + {\Omega ^{'}}}}\sqrt {\frac{x}{{{{\bar \gamma }_{1,i}}}}} \left| {\begin{array}{*{20}{c}}
				-  \\
				{\alpha ,j}
				\end{array}} \right.} \right]dx}_{{\Xi _1}} \hfill .\\
	\end{gathered}
	\end{equation}
Using [\cite{article-14}, (11) and (21)], we can solve the integral ${\Xi _1}$ as
\vspace{-0.15cm}
	\begin{equation}\label{20}
	{\Xi _1} \!\!=\!\! \frac{{{2^{\alpha  \!+\! j}}}}{{2\pi }}G_{4,1}^{1,4}\!\!\left[ \!\!{{{\!\!\left(\!\! {\frac{{{g_0}\beta  + {\Omega ^{'}}}}{{\alpha \beta }}} \!\!\right)\!\!}^2}\!\!\!\!16s{{\bar \gamma }_{1,i}}\left| \!\!\!\!{\begin{array}{*{20}{c}}
			{\frac{{2 - \alpha }}{2},\frac{{1 - \alpha }}{2},\frac{{2 - j}}{2},\frac{{1 - j}}{2}}\!\!\!\!\! \\
			0
			\end{array}} \right.} \right ].
	\end{equation}
Substituting (20) into (19), we obtain a closed-form expression for the MGF as
\vspace{-0.15cm}
\begin{equation}\label{21}
\begin{gathered}
{M_{{\gamma _{1,i}}}}\left( s \right) = \frac{A}{4}\sum\nolimits_{j = 1}^\beta  {{c_j}} \frac{{{2^{\alpha  + j}}}}{{2\pi }}   \hfill \\
\times G_{4,1}^{1,4}\left[ {{{\left( {\frac{{{g_0}\beta  + {\Omega ^{'}}}}{{\alpha \beta }}} \right)}^2}16s{{\bar \gamma }_{1,i}}\left| {\begin{array}{*{20}{c}}
		{\frac{{2 - \alpha }}{2},\frac{{1 - \alpha }}{2},\frac{{2 - j}}{2},\frac{{1 - j}}{2}} \\
		0
		\end{array}} \right.} \right] \hfill .\\
\end{gathered}
\end{equation}
\iffalse
\begin{figure*}
	\centering
	\begin{minipage}{0.32\textwidth}
		\includegraphics[width=2.5in,height=2in]{fig1.pdf}
		\caption{Beampattern for  UE }
		\label{fig:13averageLostTime}
	\end{minipage}
	\begin{minipage}{0.32\textwidth}
		\includegraphics[width=2.3in,height=2in]{fig2.pdf}
		\caption{Minimal ASR versus  the uncertain region}
		\label{fig.3 Minimal ASR versus the edge length of the uncertain region}
	\end{minipage}
\end{figure*}
\fi
According to the derived BF vectors in the proposed algorithm, the SINR in (9) for the $\emph{k}$-th user can be written as
\vspace{-0.15cm}
\begin{equation}\label{22}
\gamma _{2,k}^{} = \frac{{{{\bar \gamma }_2}{{\left| {{\rho _k}} \right|}^2}{{\left| {{{\mathbf{a}}^H}({\phi _k}){\mathbf{w}}_k^ * } \right|}^2}}}{{\sum\nolimits_{j = 1,j \ne k}^K {{{\bar \gamma }_2}{{\left| {{\rho _k}} \right|}^2}{{\left| {{{\mathbf{a}}^H}({\phi _k}){\mathbf{w}}_j^ * } \right|}^2}}  + 1}}
\end{equation}
where  ${\bar \gamma _2} = \frac{{{P_{2,k}}}}{{\sigma _{}^2}}$ is the average SNR. By defining ${X_k} \triangleq {\varphi _X}{\left| {{\rho _k}} \right|^2}$  with  ${\varphi _X} = \sum\nolimits_{j = 1}^K {{{\bar \gamma }_2}{{\left| {{{\mathbf{a}}^H}({\phi _k}){\mathbf{w}}_j^ * } \right|}^2}} $ and  ${Y_k} \triangleq {\varphi _Y}{\left| {{\rho _k}} \right|^2}$ with ${\varphi _Y} = \sum\nolimits_{j = 1,j \ne k}^K {{{\bar \gamma }_2}{{\left| {{{\mathbf{a}}^H}({\phi _k}){\mathbf{w}}_j^ * } \right|}^2}} $, with the help of (5), we obtain the cumulative distribution function of  $X^{'} \in \left\{ {{X_k},{Y_k}} \right\}$  as
\vspace{-0.15cm}
	\begin{equation}\label{23}
	\begin{split}
	{F_{X^{'}}}\left( x \right) &= 1 -{a_1}\sum\nolimits_{p = 0}^{{m_k} - 1} {\frac{{{{\left( {1 - {m_k}} \right)}_p}{{\left( { - {a_2}} \right)}^p}}}{{a_3^{p + 1}p!}}} \\
	&\times \exp \left( { - \frac{{{a_3}x}}{{{\varphi _{X^{'}}}}}} \right)\sum\nolimits_{n = 0}^p {\frac{{a_3^n}}{{\varphi _{X^{'}}^nn!}}} {x^n},\\
	\end{split}
	\end{equation}
where  ${a_1} = {\left( {2{b_k}{m_k}/2{b_k}{m_k} + {\Omega _k}} \right)^{{m_k}}}/2{b_k}$, 

\noindent ${a_2} = {\Omega _k}/\left( {2{b_k}\left( {2{b_k}{m_k} + {\Omega _k}} \right)} \right)$ and ${a_3} = 1/(2{b_k}) - {a_2}.$
\vspace{-0.2cm} 
\subsection{Ergodic Capacity}
To calculate the ergodic capacity in (10) analytically, we derive expressions of ${C_1}$  and  ${C_2}$ separately. First,  we can express  ${C_1}$  in terms of MGF as [\cite{6152071-15}]
\vspace{-0.15cm}
	\begin{equation}\label{24}
	\setlength{\abovedisplayskip}{3pt}
	\setlength{\belowdisplayskip}{3pt}
	\begin{gathered}
	{C_1} = \frac{1}{{\ln 2}}\sum\nolimits_{t = 1}^T {{V_t}}  \times \varphi \left( {{S_t}} \right)   \hfill \\
\times	\left\{ {{{\left. {\left[ {M_{{\gamma _{1,1}}}^{}\left( s \right)M_{{\gamma _{1,2}}}^{\left( 1 \right)}\left( s \right) + M_{{\gamma _{1,2}}}^{}\left( s \right)M_{{\gamma _{1,1}}}^{\left( 1 \right)}\left( s \right)} \right]} \right|}_{s \to {S_t}}}} \right\} \hfill \\
	\end{gathered}
	\end{equation}
where ${V_t} = {\pi ^2}\sin \left( {\frac{{2t - 1}}{{2T}}\pi } \right)/4T{\cos ^2}\left( {\frac{\pi }{4}\cos \left( {\frac{{2t - 1}}{{2T}}\pi } \right) + \frac{\pi }{4}} \right)$, ${S_t} = \tan \left( {\frac{\pi }{4}\cos \left( {\frac{{2t - 1}}{{2T}}\pi } \right) + \frac{\pi }{4}} \right)$, $\varphi \left( {{S_t}} \right) =  - G_{2,1}^{0,2}\left[ {\frac{1}{{{S_t}}}\left| {\begin{array}{*{20}{c}}
	{1,1} \\
	0
	\end{array}} \right.} \right]$. Besides,  $M_{{\gamma _{1,i}}}^{\left( 1 \right)}\left( s \right)$ denotes the first derivative of ${M_{{\gamma _{1,i}}}}\left( s \right),$  which can be computed by
\vspace{-0.15cm}
\begin{equation}\label{25}
\begin{gathered}
M_{{\gamma _{1,i}}}^{\left( 1 \right)}\left( s \right) =  - \frac{A}{4}\sum\nolimits_{j = 1}^\beta  {{c_j}} \frac{{{2^{\alpha  + j}}{s^{ - 1}}}}{{2\pi }}  \hfill \\
\times G_{4,1}^{1,4}\left[ {{{\left( {\frac{{{g_0}\beta  + {\Omega ^{'}}}}{{\alpha \beta }}} \right)}^2}16{{\bar \gamma }_1}s\left| {\begin{array}{*{20}{c}}
		{\frac{{2 - \alpha }}{2},\frac{{1 - \alpha }}{2},\frac{{2 - j}}{2},\frac{{1 - j}}{2}} \\
		1
		\end{array}} \right.} \right] .\hfill \\
\end{gathered}
\end{equation}
 Thus, from (21), (24) and (25), we can acquire the closed-form expression for ${C_1}$.
Next, as for ${C_2} = \sum\limits_{k \in {\mathbf{U}}}^{} {E\left[ {{{\log }_2}\left( {1 + {\gamma _{2,k}}} \right)|{\gamma _{2,k}} \geqslant {\Lambda _{th}}} \right]} $, it can be calculated as
\vspace{-0.15cm}
	\begin{equation}\label{26}
\begin{aligned}
{C_2} &= \frac{1}{{\ln 2}}\sum\limits_{k \in {\bf{U}}}^{} {\left( {E\left[ {\ln \left( {1 + {X_k}} \right)|{X_k} \ge {\Lambda _{th,X}}} \right]} \right.} \\
&\left. { - E\left[ {\ln \left( {1 + {Y_k}} \right)|{Y_k} \ge {\Lambda _{th,Y}}} \right]} \right).
\end{aligned}
	\end{equation}
Using the integral 
\vspace{-0.15cm}
\begin{equation}\label{27}
   \begin{split}
&\int_{{\Lambda _{th,X'}}}^\infty  {\ln \left( {1 + X} \right)} f\left( x \right)dx = \\
&\underbrace {\left[ {1 - F\left( {{\Lambda _{th,X'}}} \right)} \right]\ln \left( {1 + {\Lambda _{th,X'}}} \right)}_{{\Xi _2}} + \underbrace {\int_{{\Lambda _{th,X'}}}^\infty  {\frac{{1 - F\left( x \right)}}{{1 + x}}} dx}_{{\Xi _3}},
  \end{split}
\end{equation}
we have
\vspace{-0.15cm}
	\begin{equation}\label{28}
	\begin{gathered}
	{\Xi _2} = {a_1}\ln \left( {1 + {\Lambda _{th,X'}}} \right)   \hfill \\
	\times\!\! \sum\limits_{p = 0}^{{m_k} - 1} {\sum\limits_{n = 0}^p {\frac{{a_3^n}}{{\varphi _{X'}^nn!}}} \Lambda _{th,X'}^n\frac{{{{\left( {1\!\! -\!\! {m_k}} \right)}_p}{{\left( { \!-\! {a_2}} \right)}^p}}}{{a_3^{p + 1}p!}}} \exp \left(\!\! { - \frac{{{a_3}{\Lambda _{th,X'}}}}{{{\varphi _{X'}}}}}\!\! \right) \hfill. \\
	\end{gathered}
	\end{equation}
Denoting  $u = x - {\Lambda _{th,X'}}$, the integral ${\Xi _3}$  can be transformed to a form as $\int_0^\infty  {\frac{{1 - F\left( {u + {\Lambda _{th,X'}}} \right)}}{{1 + u + {\Lambda _{th,X'}}}}} du$. Using (23), we can write ${\Xi _3}$ as
\vspace{-0.15cm}
	\begin{equation}\label{29}
	\begin{gathered}
	{\Xi _3} = {a_1}\sum\limits_{p = 0}^{{m_k} - 1} {\sum\limits_{n = 0}^p {\frac{{a_3^n}}{{\varphi _{X'}^nn!}}} \frac{{{{\left( {1 - {m_k}} \right)}_p}{{\left( { - {a_2}} \right)}^p}}}{{a_3^{p + 1}p!}}}    \hfill \\
\times	\exp \left( { - \frac{{{a_3}{\Lambda _{th,X'}}}}{{{\varphi _{X'}}}}} \right)\underbrace {\int_0^\infty  {\exp \left( { - \frac{{{a_3}u}}{{{\varphi _{X'}}}}} \right)\frac{{{{\left( {u + {\Lambda _{th,X'}}} \right)}^n}}}{{1 + u + {\Lambda _{th,X'}}}}} du}_{{\Xi _4}} \hfill. \\
	\end{gathered}
	\end{equation}
To simplify integral ${\Xi _4}$  in (29), using [\cite{gradshteyn2014table-16}, eq. (3.352-4) and (3.353-5)], we can obtain an expression as shown (30).
\begin{figure*}[ht]%b!
	\setcounter{equation}{29}
\begin{equation}\label{30}
{\Xi _4} \!\!=\!\! \sum\limits_{q = 0}^n \!\!{\left(\!\!\!\! {\begin{array}{*{20}{c}}
		n \\
		q
		\end{array}} \!\!\!\!\right)} \Lambda _{th,X'}^{n - q} \!\!\times \!\! \hfill 
\left\{ \begin{gathered}
{\left( { \!-\! 1} \right)^{q \!- \!1}}{\left( {1\! +\! {\Lambda _{th,X'}}} \right)^q}{e^{\frac{{{a_3}\left( {1 + {\Lambda _{th,X'}}} \right)}}{{{\varphi _{X'}}}}}}{\text{Ei}}\left(\!\! {\! -\! \frac{{{a_3}\left( {1 \!+\! {\Lambda _{th,X'}}} \!\!\right)}}{{{\varphi _{X'}}}}} \right) \hfill  \\
\!+\! \sum\limits_{l = 1}^q {\left( {l\! -\! 1} \right)!{{\left[ {\! -\! \left( {1\! +\! {\Lambda _{th,X'}}} \right)} \right]}^{q - l}}{{\left( {\frac{{{a_3}}}{{{\varphi _{X'}}}}} \right)}^{ - l}},q > 0}  \hfill \\
\begin{array}{*{20}{c}}
{ - \exp \left( {\frac{{{a_3}\left( {1 + {\Lambda _{th,X'}}} \right)}}{{{\varphi _{X'}}}}} \right){\text{Ei}}\left( { - \frac{{{a_3}\left( {1 + {\Lambda _{th,X'}}} \right)}}{{{\varphi _{X'}}}}} \right)},&{q = 0}
\end{array} \hfill \\
\end{gathered}  \right. \hfill \\
\end{equation}
%	\noindent \rule{\textwidth}{.5pt}%\vskip3pt
\end{figure*}
From (26)-(30) and $X^{'} \in \left\{ {{X_k},{Y_k}} \right\}$, we can obtain a closed-form expression of ${C_2}$. Finally, applying the derived expressions for ${C_1}$  and ${C_2}$  to (10), we obtain an analytical EC expression, which is omitted due to space limitation.
\begin{figure}[htbp!]
		\vspace{-0.45cm}  %调整图片与上文的垂直距离
		\setlength{\belowcaptionskip}{-0.4cm}   %调整图片标题与下文距离
	\centering
	\includegraphics[width=7.6cm]{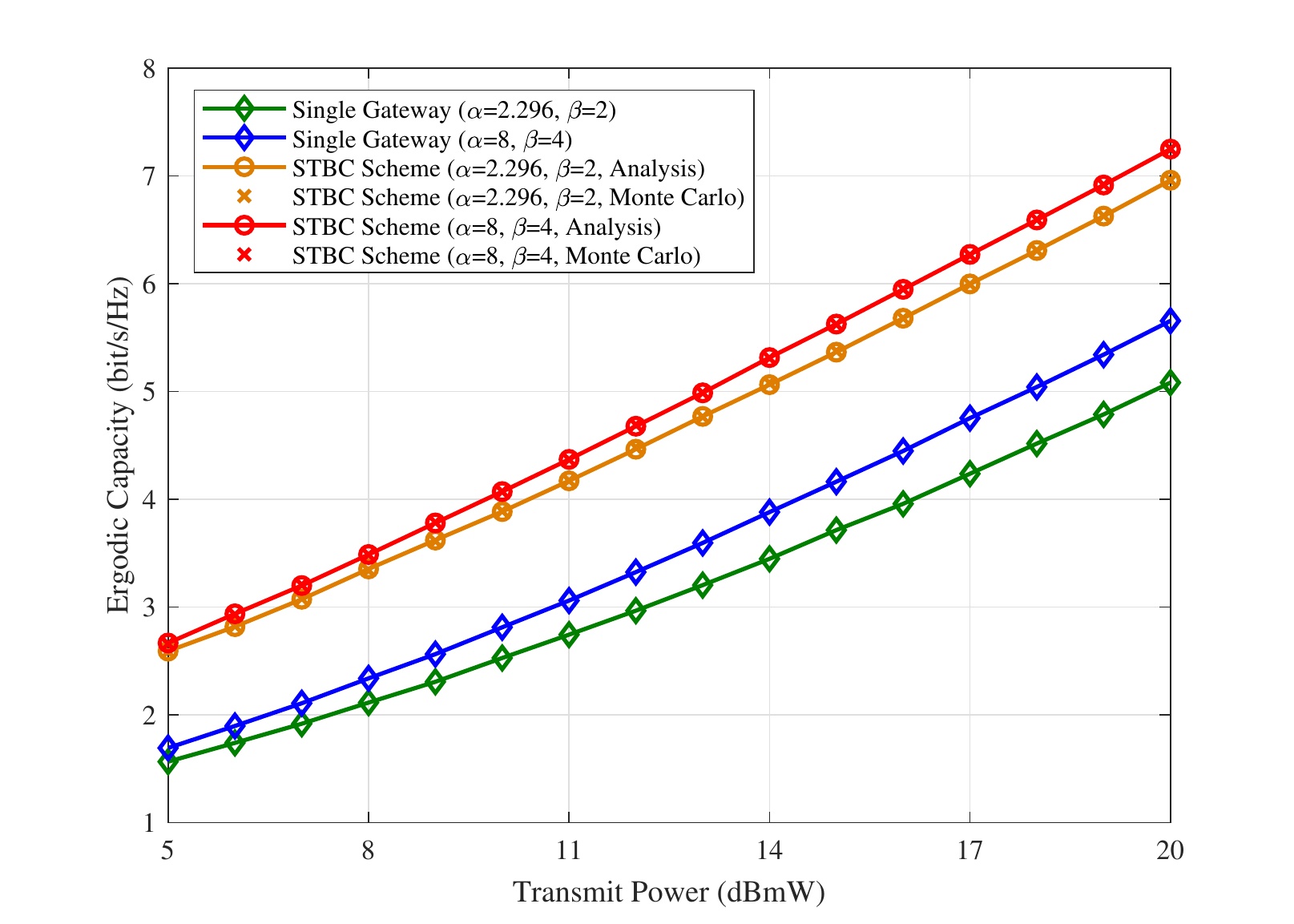}
	\caption{Ergodic capacity versus transmit power under different turbulence conditions.}\label{Fig1}
\end{figure}
\vspace{-0.2cm}
\section{Numerical Results}
This section provides numerical results to confirm the correctness of our theoretical analysis and the superiority of the proposed schemes. We take the ZF scheme [\cite{9152013-7}], [\cite{9108615-8}], and the signal-to-leakage-and-noise ratio (SLNR) scheme for the user link, as benchmark schemes. Note that the SLNR is defined as a special case of virtual SINR when the weight coefficient ${\mu _{k,j}=1}$, which is not optimized. In addition, typical simulation parameters were taken from the existing works  [\cite{9108615-8}], [\cite{7859265-10}].
%\vspace{-0.4cm}

%\vspace{-0.9cm}
\begin{figure}[htbp!]
	\vspace{-0.45cm}  %调整图片与上文的垂直距离
	\setlength{\belowcaptionskip}{-0.4cm}   %调整图片标题与下文距离
	\centering
	\includegraphics[width=7.6cm]{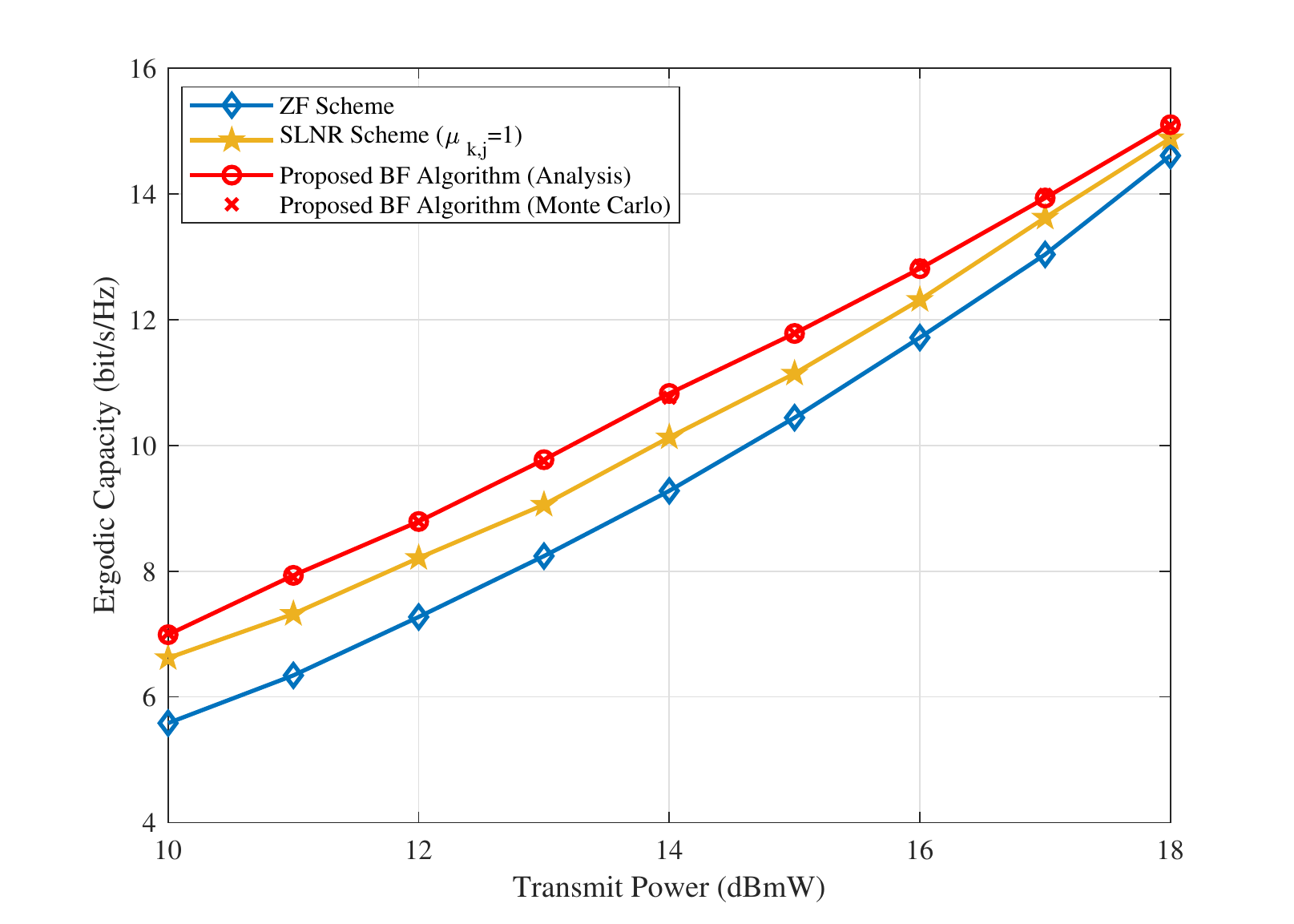}
	\caption{Ergodic capacity versus transmit power with different BF schemes.}\label{Fig2}
\end{figure}

%\vspace{-0.02cm}
Figures 1 and 2 depict the ergodic capacity of the considered system along with other reference schemes. As expected, the analytical results match well with Monte Carlo simulations, which verify the correctness of theoretical analysis for the proposed STBC scheme in the feeder link and the proposed BF algorithm in the user link. From Fig. 1, the lower atmospheric turbulence effects can result in better performance of both STBC scheme and single gateway. Moreover, the EC of STBC scheme improves significantly over the single-gateway scenario. Moreover, we derived a closed-form EC expression for such a system.
In Fig. 2, it can be seen that the proposed BF algorithm outperforms the benchmark schemes. This is because that the weight coefficient ${\mu _{k,j}}$ has further been optimized iteratively compared with the SLNR scheme and the low-complexity user selection scheme considered in the proposed BF algorithm.

\vspace{-0.2cm}
\section{Conclusion}
We studied the forward link ergodic capacity of an HTS system with mixed FSO-RF transmission. In particular, the Alamouti STBC scheme was presented in the feeder link and a BF algorithm was proposed for the user link based on the average virtual SINR and one-bit feedback to maximize the EC.  	
	%The analytical expression is different from the existing works where QoS requirement for each user was not considered.}
	It is shown that STBC diversity scheme for the feeder link can mitigate the effects of turbulence and the proposed BF algorithm for the user link can achieve higher EC than the reference schemes.
In our future works, we will further improve the proposed BF scheme and evaluate the performance gap  to optimal BF schemes using the weighted minimum mean square error approach.

\appendices

\ifCLASSOPTIONcaptionsoff
  \newpage
\fi

%\scriptsize
\footnotesize
%\small
\bibliographystyle{IEEEtran} %(plain可以替换为你所投杂志的BTS文件名，其控制引用参考文献格式）
\bibliography{IEEEabrv,letter} % mybib是Endnote导出的文件名，即你的文献库bib格式

\begin{thebibliography}{10}
\providecommand{\url}[1]{#1}
\csname url@samestyle\endcsname
\providecommand{\newblock}{\relax}
\providecommand{\bibinfo}[2]{#2}
\providecommand{\BIBentrySTDinterwordspacing}{\spaceskip=0pt\relax}
\providecommand{\BIBentryALTinterwordstretchfactor}{4}
\providecommand{\BIBentryALTinterwordspacing}{\spaceskip=\fontdimen2\font plus
\BIBentryALTinterwordstretchfactor\fontdimen3\font minus
  \fontdimen4\font\relax}
\providecommand{\BIBforeignlanguage}[2]{{%
\expandafter\ifx\csname l@#1\endcsname\relax
\typeout{** WARNING: IEEEtran.bst: No hyphenation pattern has been}%
\typeout{** loaded for the language `#1'. Using the pattern for}%
\typeout{** the default language instead.}%
\else
\language=\csname l@#1\endcsname
\fi
#2}}
\providecommand{\BIBdecl}{\relax}
\BIBdecl

\bibitem{8746876-1}
A.~I. {Perez-Neira}, M.~A. {Vazquez}, M.~R.~B. {Shankar}, S.~{Maleki}, and
  S.~{Chatzinotas}, ``Signal processing for high-throughput satellites:
  Challenges in new interference-limited scenarios,'' \emph{{IEEE} Signal
  Process. Mag.}, vol.~36, no.~4, pp. 112--131, Jul. 2019.

\bibitem{8894851-2}
Q.~{Huang}, M.~{Lin}, W.-P. {Zhu}, S.~{Chatzinotas}, and M.-S. {Alouini},
  ``Performance analysis of integrated satellite-terrestrial multiantenna relay
  networks with multiuser scheduling,'' \emph{{IEEE} Trans. Aerosp. Electron.
  Syst.}, vol.~56, no.~4, pp. 2718--2731, Aug. 2020.

\bibitem{9205852-3}
Z.~Lin, M.~Lin, B.~Champagne, W.-P. Zhu, and N.~Al-Dhahir, ``Secure and energy
  efficient transmission for {RSMA}-based cognitive satellite-terrestrial
  networks,'' \emph{{IEEE} Wireless Commun. Lett.}, vol.~10, no.~2, pp.
  251--255, 2021.

\bibitem{7553489-4}
H.~{Kaushal} and G.~{Kaddoum}, ``Optical communication in space: Challenges and
  mitigation techniques,'' \emph{{IEEE} Commun. Surveys Tuts.}, vol.~19, no.~1,
  pp. 57--96, 1st Quart. 2017.

\bibitem{8739771-6}
N.~K. {Lyras}, C.~I. {Kourogiorgas}, T.~T. {Kapsis}, and A.~D. {Panagopoulos},
  ``Ground-to-satellite optical link turbulence effects: Propagation modelling
  transmit diversity performance,'' in \emph{Proc. 13th European Conference on
  Antennas and Propagation (EuCAP)}, Apr. 2019, pp. 1--5.

\bibitem{9152013-7}
I.~{Ahmad}, K.~D. {Nguyen}, N.~{Letzepis}, and G.~{Lechner}, ``On the
  next-generation high throughput satellite systems with optical feeder
  links,'' \emph{{IEEE} Syst. J.}, pp. 1--12, Jul. 2020.

\bibitem{9108615-8}
E.~{Zedini}, A.~{Kammoun}, and M.-S. {Alouini}, ``Performance of multibeam very
  high throughput satellite systems based on {FSO} feeder links with {HPA}
  nonlinearity,'' \emph{{IEEE} Trans. Wireless Commun.}, vol.~19, no.~9, pp.
  5908--5923, Sept. 2020.

\bibitem{kaushal2017free-9}
H.~Kaushal, V.~Jain, and S.~Kar, \emph{Free {S}pace {O}ptical
  {C}ommunication}.\hskip 1em plus 0.5em minus 0.4em\relax Springer, 2017,
  vol.~1.

\bibitem{7859265-10}
M.~J. {Saber} and S.~M.~S. {Sadough}, ``On secure free-space optical
  communications over {M}\'{a}laga turbulence channels,'' \emph{{IEEE} Wireless
  Commun. Lett.}, vol.~6, no.~2, pp. 274--277, Apr. 2017.

\bibitem{8970423-11}
H.~{Ajam}, M.~{Najafi}, V.~{Jamali}, and R.~{Schober}, ``Ergodic sum rate
  analysis of {UAV}-based relay networks with mixed {RF-FSO} channels,''
  \emph{IEEE Open J. Commun. Soc.}, vol.~1, pp. 164--178, Feb. 2020.

\bibitem{6130547-12}
S.~{Park}, H.~{Park}, H.~{Kong}, and I.~{Lee}, ``New beamforming techniques
  based on virtual {SINR} maximization for coordinated multi-cell
  transmission,'' \emph{{IEEE} Trans. Wireless Commun.}, vol.~11, no.~3, pp.
  1034--1044, Mar. 2012.

\bibitem{Mullen-13}
K.~Mullen, ``A note on the ratio of two independent random variables,''
  \emph{The Amer. Statist.}, vol.~21, no.~3, pp. 30--31, Jun. 1967.

\bibitem{article-14}
V.~Adamchik and O.~Marichev, ``The algorithm for calculating integrals of
  hypergeometric type functions and its realization in reduce system,''
  \emph{Int. Symp. Symbolic Algebraic Comput.}, pp. 212--224, Jul. 1990.

\bibitem{6152071-15}
F.~{Yilmaz} and M.-S. {Alouini}, ``A unified {MGF}-based capacity analysis of
  diversity combiners over generalized fading channels,'' \emph{{IEEE} Trans.
  Commun.}, vol.~60, no.~3, pp. 862--875, Mar. 2012.

\bibitem{gradshteyn2014table-16}
I.~S. Gradshteyn and I.~M. Ryzhik, \emph{Table of Integrals, Series, and
  Products}.\hskip 1em plus 0.5em minus 0.4em\relax Academic Press, 2014.

\end{thebibliography}
%\bibliography{abbrvnat}

\end{document}